\documentclass[acmsmall,screen,nonacm=true, authorversion=true]{acmart}

\usepackage{amsmath}
\usepackage{xspace}
\usepackage{booktabs}
\usepackage{multirow}
\usepackage{algorithm}
\usepackage{algorithmic}

\AtEndPreamble{%
  \theoremstyle{acmdefinition}%
  \newtheorem{remark}[theorem]{Remark}%
}

\newcommand{\Fulcrum}{\textsc{Fulcrum}\xspace}
\newcommand{\KL}{D_{\mathrm{KL}}}
\newcommand{\Reg}{\mathcal{R}}

\begin{document}

\title{Topology-Aware Differential Privacy in Hierarchical Federated Learning}

\author{Murtaza Rangwala}
\orcid{0000-0000-0000-0000}
\affiliation{%
  \institution{The University of Melbourne}
  \department{School of Computing and Information Systems}
  \city{Melbourne}
  \country{Australia}}
\email{mrangwala@student.unimelb.edu.au}

\author{Richard O. Sinnott}
\orcid{0000-0000-0000-0000}
\affiliation{%
  \institution{The University of Melbourne}
  \department{School of Computing and Information Systems}
  \city{Melbourne}
  \country{Australia}}
\email{rsinnott@unimelb.edu.au}

\author{Rajkumar Buyya}
\orcid{0000-0000-0000-0000}
\affiliation{%
  \institution{The University of Melbourne}
  \department{School of Computing and Information Systems}
  \city{Melbourne}
  \country{Australia}}
\email{rbuyya@unimelb.edu.au}

\renewcommand{\shortauthors}{Rangwala et al.}

\begin{abstract}
Hierarchical federated learning places regional aggregators between clients and the
cloud, so a participant's update is observed only alongside its
neighbours'. The concealment this arrangement provides depends on the size of the
aggregation region, and regions in operational deployments vary widely.
Prevailing practice applies a single noise multiplier to every participant, calibrated
for the most exposed region, so every other participant carries more
noise than its own exposure requires. We show that this allocation problem admits an
explicit solution. We first give a silo-level differential privacy guarantee for the
mechanism, then bound the mutual information between a participant's local class
distribution and any estimate an observer positioned above the regional tier could form
of it, using an adjacency notion matched to the quantity being protected. Minimising the
worst-case bound under a fixed utility budget yields a min-max optimal allocation, which
we call \Fulcrum. The budget it recovers has a closed form we term the exposure
dispersion, a measure of how unevenly aggregation weight is concentrated within regions
relative to the most exposed one. Because this quantity follows from the region
structure and the aggregation weights alone, a practitioner can evaluate it before
training begins, and it vanishes precisely when all regions are equally exposed. On
image and text classification at $\varepsilon = 0.99$, accuracy at a matched worst-case
per-client guarantee improves by up to $14.84$ and $12.16$ percentage points where the
dispersion is large, and is exactly zero on a balanced control for which the theory
predicts parity.
\end{abstract}



\maketitle

\section{Introduction}
\label{sec:intro}

Machine learning increasingly draws on data that is distributed across
organisational boundaries and subject to rules that restrict its movement. Hospitals
hold patient records, banks hold transaction histories, and telecommunications
operators hold subscriber behaviour, and each of these is typically governed by
regulation that limits transfer to a central repository. Federated learning
\cite{mcmahan2017fedavg} addresses this constraint by keeping data in place and
exchanging model updates instead. It has since been deployed at scale in mobile
keyboard prediction \cite{hard2018keyboard}, clinical research consortia
\cite{sheller2020federated,dayan2021federated}, and cross-pharmaceutical
collaborations \cite{heyndrickx2023melloddy}.

This arrangement keeps raw data in place, but the privacy guarantee it offers is
weaker than it first appears. Model updates carry information about the data that
produced them, and a substantial body of work demonstrates how much of that
information can be recovered. Gradient inversion reconstructs training images from
observed updates \cite{zhu2019dlg,geiping2020inverting}, membership inference
determines whether a given record participated in training
\cite{shokri2017membership}, and property inference recovers aggregate attributes
of a participant's dataset from the sequence of updates it contributes
\cite{melis2019exploiting}. This last category is the concern of the present work,
because the quantity it recovers is a property of an organisation rather than of an
individual. The prevalence of a diagnosis at a clinical site, or the default rate
in a lender's portfolio, is commercially and legally sensitive in its own right.

Against this family of threats, differential privacy
\cite{dwork2006calibrating,dwork2014algorithmic} provides the standard defence.
Differentially private stochastic gradient descent \cite{abadi2016dpsgd} bounds the
influence any single contribution exerts on a released update by clipping the
contribution and then adding calibrated noise, and its federated instantiation
\cite{mcmahan2018dprnn} applies the same construction at client granularity.
Accounting across training rounds is performed with R\'enyi differential privacy
\cite{mironov2017rdp} or the moments accountant \cite{abadi2016dpsgd}, and mature
tooling makes the mechanism straightforward to deploy \cite{yousefpour2021opacus}.

These techniques are typically presented for a flat topology, yet realistic
deployments are rarely organised around a single server that serves every
participant directly. Bandwidth, latency, and administrative boundaries generally
favour an intermediate tier, in which participants report to a regional aggregator
that combines the updates it receives and forwards only the combination
\cite{liu2020clientedgecloud,abad2020hierarchical}. Cellular deployments site these
aggregators at base stations, clinical consortia site them at hospital or national
coordinators, and industrial deployments site them at regional gateways. This
arrangement is adopted for systems reasons, but it also changes the privacy
situation, since an observer positioned above the aggregation tier never sees an
individual participant's update in isolation.

\begin{figure}[t]
\centering
\includegraphics[width=0.60\columnwidth]{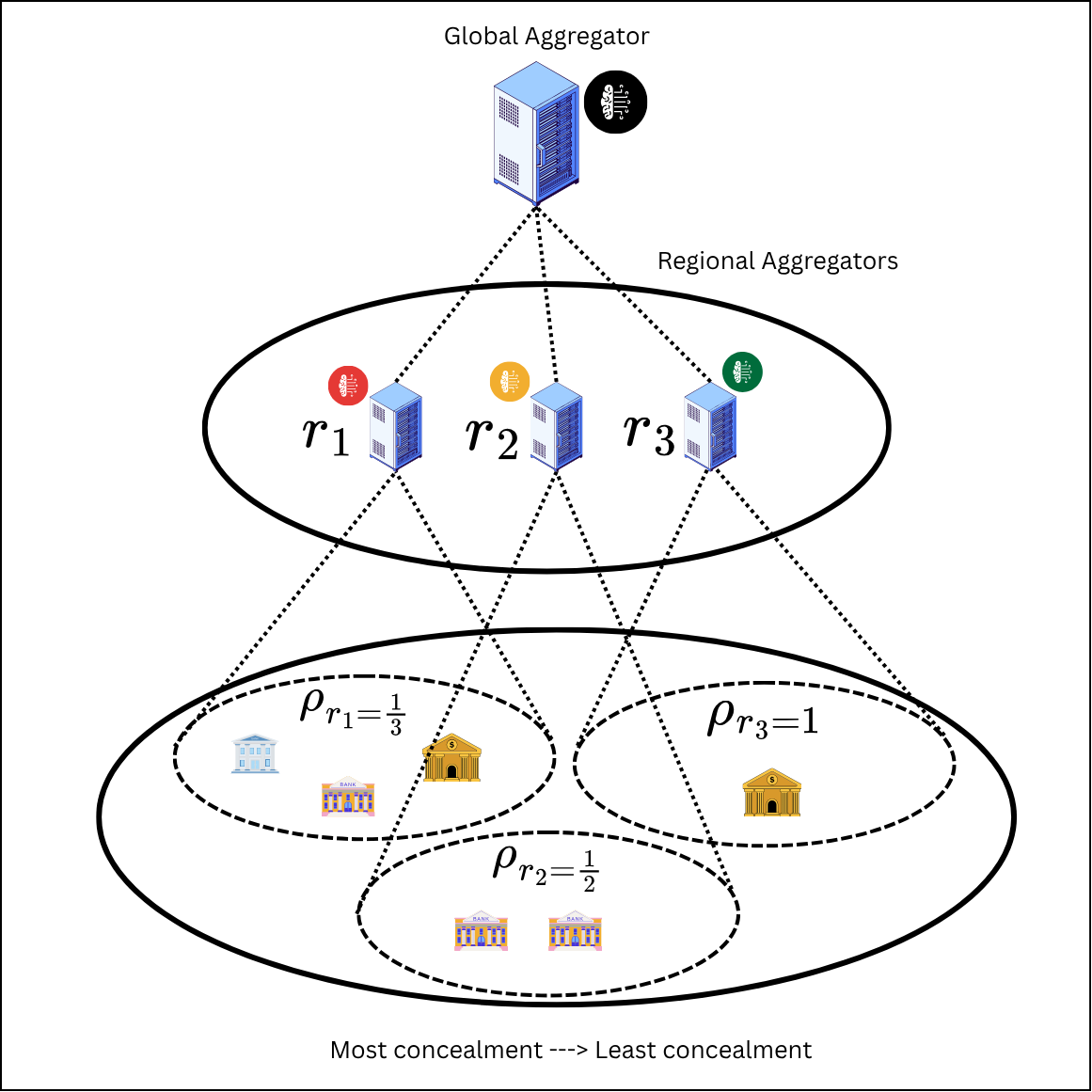}
\caption{Hierarchical federated learning with unequal regions, shown for a
consortium of banks grouped by country. Silos report to a regional aggregator,
which forwards only the combined update to the global aggregator, so that an
observer above the regional tier sees no silo in isolation. Each silo clips and
perturbs its own update before transmission, so the guarantee does not rely on the
aggregator to perturb anything. Concealment grows with the effective size of a
silo's region, measured by the exposure ratio defined in
Section~\ref{sec:model}: region $r_1$ pools three silos and conceals the most,
whereas $r_3$ holds a single silo and conceals the least. A shared noise multiplier
must satisfy $r_3$, and therefore leaves $r_1$ and $r_2$ carrying surplus noise.}
\label{fig:arch}
\end{figure}

This alteration is uneven across participants. A silo aggregated with twenty peers
is concealed by the combined noise of twenty contributions, whereas a silo sharing
its region with a single other participant receives markedly less concealment from
the structure alone. Figure~\ref{fig:arch} illustrates the arrangement, with one
region pooling three institutions and another holding only one. Operational
deployments are heterogeneous in exactly this respect \cite{kairouz2021advances}.
Cellular load distributions are long tailed \cite{wang2017cellular}, with dense
urban cells serving many devices and rural cells serving very few. Multinational
consortia contribute uneven numbers of sites per country
\cite{heyndrickx2023melloddy}, and clinical federations frequently pair one large
tertiary centre with several small clinics \cite{terrail2022flamby}.

Current practice smooths over this heterogeneity by applying a single noise
multiplier across the entire federation \cite{mcmahan2018dprnn,wei2020nbafl}. That
multiplier must satisfy the participant with the least structural concealment, so
every participant in a larger region carries more noise than its own requirement
calls for. The whole federation pays for that surplus in model accuracy, which
raises the question this paper addresses. Given a deployment's region structure,
how should noise be distributed across regions, and how much is gained by
distributing it well?

We show that this question has an explicit answer, and we call the resulting
mechanism \Fulcrum, since it balances the federation's privacy burden about the
point where every region carries an equal share of it. Our contributions are as
follows.

\begin{itemize}
\item We give two privacy results. Theorem~\ref{thm:dp} is a distribution-free
silo-level differential privacy guarantee, and is the quantity a deployment reports
as its budget. Theorem~\ref{thm:bound} bounds what an observer positioned above the
regional tier can recover about a participant's local class distribution under a
declared prior, and is the quantity the allocation minimises. Neither result is
derived from the other. Both use silo-level adjacency, matching the granularity of
the property at risk, and the bound is capped at the entropy of that property, so
whether a stated guarantee conveys anything at all is itself checkable.

\item We pose the allocation problem as minimising the worst-case value of that
bound subject to a fixed budget on the noise entering the global model, and we give
its explicit solution (Theorem~\ref{thm:allocation}). The solution is min-max
optimal and equalises the bound across regions. We also correct the utility budget,
which under weighted averaging is not the sum of the noise variances.

\item We characterise the resulting improvement in closed form
(Corollary~\ref{cor:delta}). The saving is fixed by the deployment's exposure
profile alone, a quantity we term its exposure dispersion, so a practitioner can
evaluate it before training and decide on that basis whether to adopt the method.
It is zero exactly when all regions are equally exposed, which gives an explicit
applicability test.

\item We measure the exposure dispersion for several operational structures,
finding $88\%$ for consortia organised by country and $83$ to $88\%$ for cellular
edge federations, and we validate the prediction on two modalities while reporting
achieved privacy parameters throughout. The evaluation includes a deployment for
which the theory predicts parity with the uniform baseline, together with controls
that separate exposure from non-uniformity alone.
\end{itemize}

The remainder of this paper is organised as follows. Section~\ref{sec:related}
positions the work relative to differentially private federated learning,
heterogeneous privacy budgets, hierarchical private training, and amplification
results. Section~\ref{sec:model} presents the system model, the mechanism, and the
adversary. Section~\ref{sec:theory} gives the differential privacy guarantee,
develops the per-client information bound, and discusses when that bound is
informative. Section~\ref{sec:allocation} poses and solves the allocation problem,
derives the closed form for the achievable improvement, and reports that quantity
for several operational deployments. Section~\ref{sec:eval} describes the
experimental design and results. Section~\ref{sec:discussion} discusses the scope
of the result, its assumptions, and its limitations. Section~\ref{sec:conclusion}
concludes the paper. Proofs of all formal statements appear in the appendices.

\section{Related Work}
\label{sec:related}

We position this work along four axes: private federated training, heterogeneous
privacy budgets, hierarchical and clustered private training, and amplification
arising from aggregation or topology.

\subsection{Differentially private federated learning}

Differential privacy \cite{dwork2006calibrating} bounds the effect that any single
contribution has on the distribution of a released quantity. Abadi et
al.~\cite{abadi2016dpsgd} adapted the Gaussian mechanism to deep learning by
clipping per-example gradients and perturbing their average, and introduced the
moments accountant to track cumulative loss across many training steps. McMahan et
al.~\cite{mcmahan2018dprnn} applied the same construction at client granularity
within federated averaging \cite{mcmahan2017fedavg}, clipping the client update and
thereby protecting the participation of an entire client rather than of a single
record. R\'enyi differential privacy \cite{mironov2017rdp} yields tighter
composition, and subsampling amplification \cite{wang2019subsampled} strengthens
the guarantee further when each round touches only a random subset of clients. Wei
et al.~\cite{wei2020nbafl} analysed the convergence consequences of noise addition
in this setting, and practical implementations generally follow the Opacus library
\cite{yousefpour2021opacus}. Throughout this line of work the noise multiplier is
shared by all participants, an assumption this paper revisits.

\subsection{Heterogeneous and personalised privacy budgets}

A shared multiplier can also be abandoned in a different way, by letting
participants declare individual budgets rather than deriving them from structure. A
separate line of work takes exactly this approach. Jorgensen et
al.~\cite{jorgensen2015personalized} introduced personalised differential privacy,
in which each individual declares a privacy requirement that the mechanism then
honours, and subsequent work extended the construction to federated training and to
individualised accounting \cite{boenisch2023individualized}. In these schemes the
per-participant budget enters the system as an input supplied by the user; in ours
it emerges as an output of an optimisation determined by structural position. The
two approaches are complementary, and an allocation that combines declared
preferences with structural exposure would be a natural extension, though one that
lies outside our present scope.

\subsection{Hierarchical and clustered private training}

A different source of differentiation arises not from declared preference but from
position within the topology. Hierarchical federated learning introduces
intermediate aggregators for communication efficiency and scalability
\cite{liu2020clientedgecloud, abad2020hierarchical}, and clustering methods group
participants by distributional similarity in order to train specialised models
\cite{briggs2020federated}. Differential privacy has also been applied across
tiered IoT, edge and cloud deployments in healthcare, combining noise mechanisms
with processing at the edge \cite{mangala2026dphealthcare}, although in that work
the noise is calibrated per layer rather than to the exposure each participant
receives from its own aggregation group.

Closest to our setting is the work of Chandrasekaran et
al.~\cite{chandrasekaran2022hfl}, who combine hierarchical training with
differential privacy. Clients are organised into zones with elected super-nodes,
and calibrated noise is introduced at the zone level. Their per-zone noise scale
depends on the number of participating clients in the zone, and they observe that
intermediate aggregation confers additional protection at the central aggregator.
We take the existence of this effect as our point of departure rather than as a
contribution; our addition is the optimisation. We establish that selecting the
per-region noise levels is a constrained min-max problem with an explicit solution,
and we characterise exactly how much is gained relative to uniform allocation.

The two settings differ in where the noise originates, and the difference is
quantitative. Where a trusted super-node contributes a single noise draw on behalf
of its zone, the noise required per client scales as $1/m$ in a region of $m$
equally weighted participants. Where each participant contributes its own noise
prior to transmission, as we assume here, the region's noise accumulates across $m$
independent draws and the requirement instead scales as $1/\sqrt{m}$. The factor
$\sqrt{m}$ separating these two regimes measures what a deployment pays for
declining to rely on the aggregator to inject noise on its behalf, and we state
this assumption explicitly rather than leaving it implicit in the mechanism.

Other recent work in this space likewise treats structure as given and asks how to
allocate against it, without characterising the achievable gain. Guo et
al.~\cite{guo2025fedcdp} study clustered federated learning with heterogeneous
differential privacy. Their clusters group clients of similar distribution in order
to train personalised models, and their intra-cluster weighting adjusts aggregation
weights as a logarithmic function of a client's noise level. The direction here is
the reverse of ours: they treat the noise as given and select the weights, whereas
we treat the structure as given and select the noise. Yin et
al.~\cite{yin2026hdpfedcd} allocate noise according to a per-example data-quality
metric computed within each client's local dataset; their topology is flat and
their hierarchy is internal to a client, so region size plays no part. Both works
select allocations heuristically and leave the achievable improvement
uncharacterised.

\subsection{Amplification by aggregation and by topology}

A different strand of work asks not how to allocate noise across a given
structure, but how much amplification the structure itself already provides.
Secure aggregation \cite{bonawitz2017secagg} conceals individual updates from the
server using cryptographic masking, and shuffling arguments quantify the resulting
amplification, which improves as more contributions are combined. Cyffers and
Bellet \cite{cyffers2022amplification} and Cyffers et
al.~\cite{cyffers2022muffliato} develop network differential privacy for
decentralised averaging, where the guarantee between two nodes strengthens with
their separation in the communication graph. Cuff and Yu \cite{cuff2016mi} relate
differential privacy to mutual information, furnishing the currency in which our
own bound is expressed.

This body of work computes the amplification enjoyed by a fixed and uniform
mechanism. We instead treat the amplification profile as the object to allocate
against. Where regions differ in effective size the amplification they provide is
unequal, and uniform noise is then wasteful by a quantity we compute exactly.
Table~\ref{tab:related} summarises this comparison. Making it precise requires the
system and threat model we set out next.

\begin{table}[t]
\caption{Placement relative to the closest lines of work.}
\label{tab:related}
\centering
\footnotesize
\setlength{\tabcolsep}{3.5pt}
\begin{tabular}{@{}lccc@{}}
\toprule
& Per-client & Allocation & Optimality \\
& noise varies & set by & result \\
\midrule
DP-FedAvg \cite{mcmahan2018dprnn}      & no  & shared multiplier   & no \\
Personalised DP \cite{jorgensen2015personalized} & yes & user preference & no \\
FedCDP \cite{guo2025fedcdp}            & yes & noise level         & no \\
HDP-FedCD \cite{yin2026hdpfedcd}       & yes & data quality        & no \\
Hierarchical DP \cite{chandrasekaran2022hfl} & yes & zone size     & no \\
Muffliato \cite{cyffers2022muffliato}  & no  & shared multiplier   & no \\
\textbf{This work}                     & \textbf{yes} & \textbf{region exposure} & \textbf{yes} \\
\bottomrule
\end{tabular}
\end{table}

\section{System and Threat Model}
\label{sec:model}

We now make precise the structure sketched above, namely the federation, the
mechanism run by each silo, and the observer whose knowledge the rest of the paper
bounds.

\subsection{Federation and aggregation structure}

A federation comprises $n$ silos $P_1, \dots, P_n$. Silo $i$ holds a local dataset
$D_i$ and contributes to the global model with aggregation weight $w_i = |D_i| /
\sum_j |D_j|$, so that $\sum_i w_i = 1$. Silos are partitioned into disjoint
aggregation regions, collected in $\Reg$. We write $A(i) \in \Reg$ for the region
containing silo $i$ and index regions by $r$. Training proceeds for $T$ rounds of
federated averaging \cite{mcmahan2017fedavg}. Table~\ref{tab:notation} collects the
notation used throughout.

Three properties of this structure are assumed throughout the paper. Regions are
disjoint, so each silo contributes to exactly one aggregate; membership is fixed
for the duration of training; and silo datasets are disjoint, so that no record is
held by two participants. The first two assumptions make the observer's view
decomposable region by region, which is what Section~\ref{sec:theory} relies on,
and the third makes silo-level adjacency well defined. Deployments in which a silo
feeds several aggregates, or in which membership changes between rounds, would
require the accounting to be extended, and we do not treat them here.

The region structure is a property of the deployment rather than a parameter we
select. It follows the physical network, where clients report to the base station
serving them, and it follows the organisational structure, where clinical sites
report through a national coordinator, or where member banks report through a
country-level entity as in Figure~\ref{fig:arch}.

\begin{table}[t]
\caption{Notation.}
\label{tab:notation}
\centering
\footnotesize
\setlength{\tabcolsep}{4pt}
\begin{tabular}{@{}l p{0.74\columnwidth}@{}}
\toprule
Symbol & Meaning \\
\midrule
$n$, $T$            & number of silos, number of rounds \\
$D_i$, $w_i$        & local dataset and aggregation weight of silo $i$ \\
$A(i)$, $\Reg$      & aggregation region of silo $i$, set of regions \\
$m_r$, $m_{\min}$   & number of silos in region $r$, in the smallest region \\
$C$, $\sigma_i$     & clipping bound, noise multiplier of silo $i$ \\
$Y_r^{(t)}$         & observation: the round-$t$ sum over region $r$, \eqref{eq:observation} \\
$p_i$, $R$          & sensitive-class concentration, quantisation levels \\
$H(p_i)$            & entropy of the quantised target, the vacuity ceiling \\
$\ell_i$            & lateral floor, $I(p_i; D_{-i})$ \\
$\kappa$, $\Phi_g$  & within-group coupling and group parameter of the prior \\
$m_i(\sigma)$       & per-round mechanism term, \eqref{eq:mterm} \\
$a$                 & composition coefficient, $a = 2T$ \\
$S_r$, $V_r$, $W_r$ & region noise, weight mass, and peak squared weight \\
$\rho_r$            & exposure ratio $W_r / V_r$ of region $r$ \\
$U$, $K$            & utility budget, worst-case per-client bound \\
$\delta$            & exposure dispersion, the achievable saving \\
\bottomrule
\end{tabular}
\end{table}

\subsection{Mechanism}

Given this structure, the mechanism proceeds round by round. At round $t$, silo $i$
trains locally on $D_i$ starting from the current global model $\theta^{(t-1)}$,
forms its update $\Delta_i^{(t)}$, clips the update so that $\|\Delta_i^{(t)}\| \le
C$, and adds Gaussian noise,
\begin{equation}
\tilde{\Delta}_i^{(t)} = \mathrm{clip}\big(\Delta_i^{(t)}, C\big) + \xi_i^{(t)},
\qquad
\xi_i^{(t)} \sim \mathcal{N}\!\left(0, \sigma_i^2 C^2 I\right).
\label{eq:mechanism}
\end{equation}
The regional aggregator sums the perturbed updates it receives, and the observer
sees the per-region sums
\begin{equation}
Y_r^{(t)} = \sum_{j \in r} w_j \, \tilde{\Delta}_j^{(t)} .
\label{eq:observation}
\end{equation}
Algorithm~\ref{alg:main} states the full procedure, including the allocation
derived in Section~\ref{sec:allocation}.

Three aspects of \eqref{eq:mechanism} deserve comment. Clipping applies to the
silo's entire update rather than to per-example gradients, so the mechanism
protects the participation of a whole silo, matching the granularity of the
quantity we set out to protect. Each silo contributes its own noise prior to
transmission, so the aggregator need not be relied upon to perturb anything;
Section~\ref{sec:adversary} sets out what this does and does not assume about the
aggregator itself. Finally, the amount of local computation performed between
communications leaves the privacy accounting unchanged, since each silo releases
exactly one clipped and perturbed update per round regardless of how many local
optimisation steps produced it, so the number of mechanism invocations equals $T$.
Record-level DP-SGD behaves differently in this respect, since there every local
step constitutes a separate release.

\begin{algorithm}[t]
\caption{\Fulcrum: federated training with min-max optimal regional allocation}
\label{alg:main}
\begin{algorithmic}[1]
\STATE \textbf{Input:} regions $\Reg$, weights $\{w_i\}$, budget $U$, rounds $T$, clip $C$
\STATE $V_r \leftarrow \sum_{j \in r} w_j^2$, \quad $W_r \leftarrow \max_{i \in r} w_i^2$
\STATE $K^\star \leftarrow$ root of $\sum_r 2T\,W_r/(K-\ell_r) = U$ on $(\max_r \ell_r, \infty)$
\STATE $S_r^\star \leftarrow 2T\,W_r/(K^\star - \ell_r)$ \quad for each $r \in \Reg$
\STATE $\sigma_i^2 \leftarrow S_{A(i)}^\star / V_{A(i)}$ \quad for each silo $i$
\FOR{$t = 1$ \TO $T$}
  \FOR{each silo $i$ in parallel}
    \STATE $\Delta_i^{(t)} \leftarrow \textsc{LocalTrain}(\theta^{(t-1)}, D_i)$
    \STATE $\tilde{\Delta}_i^{(t)} \leftarrow \mathrm{clip}(\Delta_i^{(t)}, C) + \mathcal{N}(0, \sigma_i^2 C^2 I)$
  \ENDFOR
  \STATE $Y_r^{(t)} \leftarrow \sum_{j \in r} w_j \tilde{\Delta}_j^{(t)}$ \quad for each $r$
  \STATE $\theta^{(t)} \leftarrow \theta^{(t-1)} + \sum_{r} Y_r^{(t)}$
\ENDFOR
\RETURN $\theta^{(T)}$
\end{algorithmic}
\end{algorithm}

\subsection{Adversary}
\label{sec:adversary}

What can an observer positioned above the aggregation tier learn from what it
sees? We consider a passive observer that follows the protocol and examines only
what its position exposes. It knows the region structure, the aggregation weights,
and the noise scales $\{\sigma_j\}$, and it observes the per-region sums
$Y_r^{(t)}$ across all rounds. Individual silo updates remain hidden from it, which
is precisely the property the aggregation tier provides. This model is standard in
the property-inference literature \cite{melis2019exploiting} and reflects an
infrastructure operator or a curious coordinator.

The position of the observer is a substantive assumption rather than a
presentational one, because the concealment the allocation exploits is supplied by
the regional sum. A regional aggregator observes the individual updates of its own
members, so a silo facing a curious aggregator is protected by $\sigma_i$ alone,
with no contribution from its neighbours. The allocation lowers $\sigma_i$ in the
larger regions, which is where it recovers budget, so it weakens that guarantee
exactly where it strengthens the regional one; Section~\ref{sec:discussion}
quantifies this effect. We therefore assume that regional aggregators are trusted
to see what they aggregate, while assuming nothing about the central aggregator or
any observer above the regional tier, and while requiring no party to inject noise
on a silo's behalf.

The observer's objective is to infer, for a target silo $i$, the sensitive-class
concentration
\begin{equation}
p_i = \frac{\big|\{x \in D_i : x \text{ belongs to the sensitive class}\}\big|}{|D_i|},
\end{equation}
quantised to $R$ levels so that $H(p_i)$ is finite. Quantisation reflects the
operational reality that an estimate is useful only at some finite resolution, and
$R$ is therefore a property of the inference the observer is attempting rather than
a free parameter of the model. Two considerations fix it. A prevalence estimate is
normally acted on at a resolution of a few percentage points, and a silo holding
$|D_i|$ records carries a sampling error of order $\sqrt{1/4|D_i|}$ on its own
concentration, so resolutions finer than that are not identifiable from the data in
any case. At the scale of our experiments, the second consideration caps the
useful resolution near $R = 28$; we adopt the coarser $R = 20$, a resolution of
$0.05$, throughout. Section~\ref{sec:eval} reports the informativeness test across
a range of $R$, since $H(p_i) \le \ln R$ means a finer quantisation would otherwise
inflate the cap for free. This target captures information organisations are
frequently obliged to protect, such as the prevalence of a condition at a clinical
site, and it is a property of the dataset rather than of any individual record.

\section{Privacy Analysis}
\label{sec:theory}

This section establishes two results about the mechanism of Section~\ref{sec:model}.
It is worth separating them at the outset, because they answer different
questions and rest on different assumptions.

The first is a differential privacy guarantee. It concerns the participation of a
silo as such, holds for every dataset and every prior, and is the quantity a
deployment reports as its $(\varepsilon, \delta_{\mathrm{DP}})$ budget. The second
bounds how much an observer learns about one designated quantity, the
sensitive-class concentration $p_i$, and is stated relative to a declared
generative model, because part of that quantity is already fixed by the other
silos before the mechanism acts at all. The allocation of
Section~\ref{sec:allocation} minimises the second; the first is computed from the
resulting noise scales and reported alongside it. Neither result is derived from
the other.

\subsection{Adjacency and the two observers}

Both results use \emph{silo-level} adjacency, under which two deployments are
adjacent when one silo's dataset $D_i$ is replaced in its entirety. This is the
granularity at which the protected quantity lives, since $p_i$ is a property of
$D_i$ as a whole and is not controlled by a guarantee over individual records.
Because each transmitted update is clipped to $\|\Delta\| \le C$ before
perturbation, replacing $D_i$ moves that update by at most $2C$, however different
the two datasets are.

Who observes matters as much as what is protected, and the mechanism admits two
distinct positions. An observer \emph{above} the regional tier sees the sums
$Y_r^{(t)}$ and never an individual update. Silo $i$'s contribution enters
$Y_{A(i)}^{(t)}$ scaled by $w_i$ and is masked by the noise of every silo in the
region, so its effective noise multiplier is
\begin{equation}
\sigma^{\mathrm{ab}}_i \;=\; \frac{\sqrt{S_r}}{w_i},
\qquad S_r \;=\; \sum_{j \in r} w_j^2 \sigma_j^2 .
\label{eq:sigmaeff}
\end{equation}
An observer \emph{within} the region, of which the regional aggregator is the
natural instance, sees $\tilde{\Delta}_i^{(t)}$ directly; no pooling applies, and
the effective multiplier is $\sigma_i$. We carry both through the analysis rather
than only the first, because the allocation moves them in opposite directions and
reporting only one would misstate what is gained.

\subsection{Differential privacy of the mechanism}

The effective multiplier just defined for each observer yields the guarantee
directly.

\begin{theorem}[Silo-level differential privacy]
\label{thm:dp}
Fix $\delta_{\mathrm{DP}} \in (0,1)$ and let $\sigma^{\ast}_i$ be the effective
multiplier for the observer under consideration, namely $\sigma^{\mathrm{ab}}_i$
of \eqref{eq:sigmaeff} above the regional tier and $\sigma_i$ within the region.
Over $T$ rounds the mechanism satisfies $(\varepsilon_i,
\delta_{\mathrm{DP}})$-differential privacy at silo-level adjacency against that
observer, with
\begin{equation}
\varepsilon_i \;=\; \inf_{\alpha > 1}
\left[\, \frac{2 T \alpha}{(\sigma^{\ast}_i)^2}
\;+\; \log\!\Big(\tfrac{\alpha - 1}{\alpha}\Big)
\;-\; \frac{\log \delta_{\mathrm{DP}} + \log \alpha}{\alpha - 1} \,\right].
\label{eq:dpeps}
\end{equation}
\end{theorem}

\begin{proof}
One release perturbs a quantity of $\ell_2$ sensitivity $2C$ with Gaussian noise of
scale $\sigma^{\ast}_i C$. The Gaussian mechanism with sensitivity $\Delta$ and
scale $s$ satisfies $(\alpha, \alpha \Delta^2 / 2s^2)$-R\'enyi differential privacy
\cite{mironov2017rdp}, which here evaluates to $\alpha (2C)^2 / (2 (\sigma^{\ast}_i
C)^2) = 2\alpha/(\sigma^{\ast}_i)^2$ per round. R\'enyi differential privacy
composes additively over the $T$ rounds, giving $2 T \alpha/(\sigma^{\ast}_i)^2$
overall, and \eqref{eq:dpeps} follows from the tightened R\'enyi-to-$(\varepsilon,
\delta_{\mathrm{DP}})$ conversion of Canonne et al.\ \cite{canonne2020discrete},
optimised over the order $\alpha$.
\end{proof}

Three points of scope apply here. The guarantee is distribution-free, so that no
property of the data and no prior enters \eqref{eq:dpeps}. It counts $T$ releases
rather than $T$ times the number of local optimisation steps, because a silo
transmits one clipped and perturbed update per round however much local
computation produced it. And it claims no subsampling amplification, since every
silo participates in every round. Values of $\varepsilon$ reported in
Section~\ref{sec:eval} are obtained from \eqref{eq:dpeps} by one-dimensional
search over $\alpha$ at $\delta_{\mathrm{DP}} = 10^{-5}$.

\subsection{The lateral floor}

An observer interested in $p_i$ draws on two sources of information: what
neighbouring silos already reveal, and what the mechanism itself leaks. We bound
each in turn. The first is available without reference to any model update at all,
because silos within an organisational group tend to have correlated
distributions. We write this floor as
\begin{equation}
\ell_i = I\big(p_i ; D_{-i}\big),
\end{equation}
evaluated under the deployment's stated generative model, where $D_{-i}$ denotes
the datasets of the remaining silos. This term flows around the mechanism rather
than through it, so noise added to silo $i$'s updates leaves it unchanged.

\begin{figure}[t]
\centering
\includegraphics{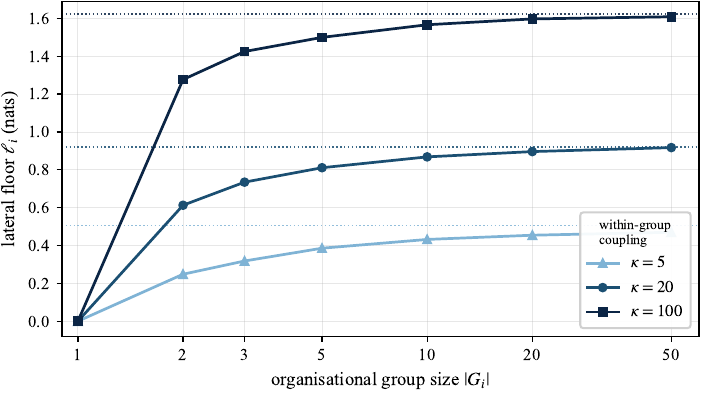}
\caption{The lateral floor $\ell_i$ as a function of organisational group size,
under a hierarchical Beta model with within-group coupling $\kappa$. The quantity
saturates at $I(p_i;\Phi_g)$, shown dotted, so it varies modestly across silos and
remains well below the entropy of the target.}
\label{fig:leverage}
\end{figure}

The floor can be computed, not merely defined. Consider a hierarchical model in
which a group-level parameter $\Phi_g$ is drawn first and each member's
concentration is drawn conditionally on it. There $\ell_i$ depends on the number of
other members of $i$'s group, increasing with that number and saturating at $I(p_i
; \Phi_g)$, which lies strictly below $H(p_i)$. Figure~\ref{fig:leverage} shows
this behaviour. Saturation is rapid: the first group member contributes roughly two
thirds of the eventual total, and the difference between a group of twenty and a
group of fifty is slight. The floor is nonetheless not constant across regions,
whereas Corollary~\ref{cor:delta} assumes that it is, and
Section~\ref{sec:allocation} quantifies what that assumption costs for each
deployment we report.

Evaluating $\ell_i$ under a stated prior, rather than as a supremum over all priors
consistent with the structure, is what makes it informative. Under a supremum one
could always select a prior that makes $p_i$ a deterministic function of a
neighbour's data, giving $\ell_i = H(p_i)$ for every silo with at least one
neighbour. The quantity would then be constant across silos, and any allocation
derived from it would coincide with the uniform one.

\subsection{The information bound}

With the lateral floor characterised, the second result follows directly. It is
the objective the allocation minimises.

\begin{theorem}[Per-client information bound]
\label{thm:bound}
Let $\hat{p}_i = f(Y)$ be an estimate formed by a deterministic function of the
observations $Y = \{Y_r^{(t)}\}$. Then
\begin{equation}
I\big(p_i ; \hat{p}_i \,\big|\, \{\sigma_j\}\big)
\le
\min\Big\{\, H(p_i), \;\; T \cdot m_i(\sigma) + \ell_i \,\Big\},
\label{eq:bound}
\end{equation}
where the mechanism term is
\begin{equation}
m_i(\sigma) = \frac{2\,w_i^2}{\sum_{j \in A(i)} w_j^2 \sigma_j^2}.
\label{eq:mterm}
\end{equation}
\end{theorem}

Appendix~\ref{app:bound} gives the proof, whose structure we summarise here. The
data processing inequality first reduces the estimate to the observations. A chain
rule then separates what the observations reveal beyond $D_{-i}$ from what
$D_{-i}$ reveals on its own, the latter being $\ell_i$ by definition. Conditioning
on the observations from earlier rounds isolates silo $i$'s contribution within
each round, since given the past, every other silo's update is a deterministic
function of the previous global model and its own data, perturbed only by
independent noise. The per-round contribution is then bounded by the largest
Kullback-Leibler divergence the mechanism produces between two silo-adjacent
inputs, and these bounds accumulate additively across rounds.

Three properties of \eqref{eq:bound} warrant emphasis.

\begin{remark}[Concealment is regional]
\label{rem:regional}
The mechanism term depends on the total noise present in the region rather than on
silo $i$'s noise in isolation. This is the formal counterpart of the observation
that a participant is concealed by its neighbours. For $m$ equally weighted silos
sharing a region and using a common scale $\sigma$, the term reduces to $2/(m
\sigma^2)$, so the noise each silo contributes to reach a given level falls as
$1/\sqrt{m}$.
\end{remark}

\begin{remark}[Structure is actionable]
Because the region enters the mechanism term, the structure is something the
allocation can act upon. This is what makes the problem of
Section~\ref{sec:allocation} non-trivial, and it distinguishes the present setting
from flat federated learning, where each silo is observed individually.
\end{remark}

\begin{remark}[Informativeness is checkable]
\label{rem:informative}
A further property concerns not what the bound says about structure, but whether
it says anything at all. The cap at $H(p_i)$ gives a test of whether a guarantee
conveys anything. The bound is informative when $T \cdot m_i(\sigma) < H(p_i) -
\ell_i$. We refer to the right-hand side as the margin, since it measures the
information about $p_i$ that the deployment structure leaves undetermined and for
which the mechanism is therefore responsible. A reported bound exceeding the
entropy of the target conveys nothing about it. Section~\ref{sec:eval} verifies the
condition at the operating point we evaluate, and across a range of quantisation
resolutions, since $H(p_i) \le \ln R$ would otherwise let a finer resolution
satisfy it for free.
\end{remark}

The bound in \eqref{eq:bound} is stated for whatever noise scales $\{\sigma_j\}$
the deployment happens to use, and nothing in its derivation requires those scales
to be equal across silos. This freedom is exactly what Section~\ref{sec:allocation}
optimises.

\section{Optimal Allocation}
\label{sec:allocation}

What remains is to determine the noise scales themselves. We first pose the
problem precisely, then derive its solution in closed form.

\subsection{Formulating the problem}

Let $S_r = \sum_{j \in r} w_j^2 \sigma_j^2$ denote the noise variance region $r$
contributes to the aggregate. The utility cost of privacy is the total noise
variance entering the global model,
\begin{equation}
U = \sum_i w_i^2 \sigma_i^2 = \sum_{r \in \Reg} S_r ,
\label{eq:budget}
\end{equation}
in place of the unweighted sum $\sum_i \sigma_i^2$. The distinction matters under
weighted averaging, where a silo's noise enters the aggregate scaled by its
weight. Adopting the unweighted sum both misstates the cost and leaves the
optimisation degenerate, since it would then be cheapest to concentrate a region's
entire noise allowance on its highest-weight member.

Writing $W_r = \max_{i \in r} w_i^2$ for the peak squared weight in region $r$, the
worst-case bound within the region is $a W_r / S_r + \ell_r$ with $a = 2T$. The
allocation problem is therefore
\begin{equation}
\min_{\{S_r > 0\}} \; \max_{r \in \Reg} \left[ \frac{a W_r}{S_r} + \ell_r \right]
\quad \text{subject to} \quad \sum_{r} S_r \le U .
\label{eq:opt}
\end{equation}

\subsection{Solving the allocation problem}

The problem in \eqref{eq:opt} turns out to admit an explicit solution.

\begin{theorem}[Min-max optimal allocation]
\label{thm:allocation}
For any budget $U > 0$, the solution of \eqref{eq:opt} is
\begin{equation}
S_r^{\star} = \frac{a W_r}{K^{\star} - \ell_r},
\label{eq:alloc}
\end{equation}
where $K^{\star}$ is the unique root of $g(K) = \sum_r a W_r / (K - \ell_r) = U$
on $(\max_r \ell_r, \infty)$. The achieved worst-case bound equals $K^{\star}$ and
is equalised across regions.
\end{theorem}

The objective is non-smooth, and introducing a slack variable for the maximum
converts \eqref{eq:opt} into a convex program with a linear objective. Slater's
condition holds, so the Karush-Kuhn-Tucker conditions are necessary and
sufficient, and they yield \eqref{eq:alloc} directly. Uniqueness of $K^{\star}$
follows because $g$ is continuous and strictly decreasing on the relevant interval,
diverges at its left endpoint, and vanishes at infinity. The per-region allocation
\eqref{eq:alloc} is explicit once $K^{\star}$ is known, and $K^{\star}$ itself is
the root of a scalar equation rather than an expression in the inputs, except when
the floors coincide, where it reduces to $\ell + a\sum_r W_r / U$.
Appendix~\ref{app:alloc} gives the full argument. Computing $K^\star$ by bisection
to tolerance $\epsilon$ costs $O(|\Reg| \log(1/\epsilon))$ operations and is
performed once, before training begins.

\subsection{The achievable improvement}

The optimal allocation gives each region exactly what it needs, and the size of
the gain is best seen against the uniform baseline it replaces. Uniform allocation
applies one scale $\sigma$ to every silo, giving $S_r = \sigma^2 V_r$ with $V_r =
\sum_{j \in r} w_j^2$. Reaching a target $K$ then requires satisfying the most
demanding region, which fixes $\sigma^2$ and hence the total budget. Comparing the
budgets the two schemes require to reach the same target gives the following.

\begin{corollary}[Budget saved at equal privacy]
\label{cor:delta}
Define the exposure ratio of region $r$ as
\begin{equation}
\rho_r = \frac{W_r}{V_r} = \frac{\max_{i \in r} w_i^2}{\sum_{i \in r} w_i^2},
\end{equation}
and let $\langle \rho \rangle_V = \sum_r \rho_r V_r / \sum_r V_r$. Where the
lateral floors $\ell_r$ agree across regions, the fraction of the noise budget
saved by \eqref{eq:alloc} relative to uniform allocation at the same target is
\begin{equation}
\delta = 1 - \frac{\langle \rho \rangle_V}{\rho_{\max}} .
\label{eq:delta}
\end{equation}
\end{corollary}

Appendix~\ref{app:delta} gives the derivation. One point of interpretation is
worth fixing immediately, since $\delta$ is a fraction of a \emph{variance} budget
and is easily misread as a fraction of a noise level. A saving of $\delta$
corresponds to a reduction in noise scale of $1/\sqrt{1-\delta}$, so the $88\%$
recorded below for a consortium amounts to a factor of $2.9$ in $\sigma$, and the
$79.2\%$ of the severe profile amounts to a factor of $2.2$. We quote $\delta$ in
variance terms throughout, because that is the quantity the budget constrains.

The quantity $\rho_r$ is the inverse participation ratio of the weights within
region $r$. For $m$ equally weighted silos it equals $1/m$, so it measures the
region's effective size, and it correctly accounts for a region dominated by one
large participant. Three consequences follow.

First, the dispersion $\delta$ is determined entirely by the exposure profile
$\{\rho_r\}$, which the region structure and the aggregation weights fix between
them. Region sizes alone do not determine it: two deployments partitioned
identically can differ in $\delta$ when their silos differ in size, as
Section~\ref{sec:eval} shows. What matters is that $\delta$ is independent of the
dataset, the model, the number of rounds, and the privacy target, so a
practitioner can compute it from the deployment description before training and
decide on that basis whether to adopt the method.

Second, it equals zero exactly when $\rho_r$ agrees across regions, that is, when
all regions are equally exposed. We use \emph{effective size} for $1/\rho_r$
throughout, since for equally weighted silos it equals the region's membership
count, so exposure and effective size are reciprocals of one another. The optimal
allocation then coincides with the uniform one, and we present this as an
applicability test to be stated in advance of training.

Third, flat federated learning, in which each silo forms its own region and its
update is observed individually, has $\rho_r = 1$ for every silo and therefore
$\delta = 0$ however unequal the silo sizes may be. An aggregation tier is what
creates the opportunity to reallocate, which places hierarchical deployments
within the scope of this paper.

\subsection{The cost of the equal-floor assumption}

Corollary~\ref{cor:delta} treats $\ell_r$ as common to all regions, whereas
Section~\ref{sec:theory} shows it growing with group size. Regions in the
heterogeneous profiles differ in size by an order of magnitude, so the assumption
is loosest exactly where $\delta$ is largest, and the direction of the error is
unfavourable: larger regions carry a larger floor, hence a smaller $K - \ell_r$,
hence a larger $S_r$ than the common-floor calculation implies.

Solving \eqref{eq:alloc} numerically with region-dependent $\ell_r$ taken from the
model of Figure~\ref{fig:leverage} at $\kappa = 20$ and $K = 2.2$ nats gives the
comparison in Table~\ref{tab:ellgap}. The common-floor value overstates the
attainable saving by $0.6$ to $4.8$ percentage points, with the largest error on
the most heterogeneous profile. The ordering of deployments is preserved, so the
common-floor expression remains serviceable for ranking structures and for the
$\delta = 0$ applicability test, and the numerical solution should be preferred
whenever the saving itself is being quoted.

\begin{table}[t]
\caption{Saving attainable under the common-floor assumption of Corollary~\ref{cor:delta}, against a numerical solution carrying region-dependent $\ell_r$ ($\kappa = 20$, $K = 2.2$ nats). Region profiles are instantiated at the evaluated federation size $n = 96$, since $\ell_r$ depends on the absolute number of silos sharing a region while $\delta$ does not.}
\label{tab:ellgap}
\centering
\footnotesize
\setlength{\tabcolsep}{5pt}
\begin{tabular}{@{}lccc@{}}
\toprule
& & \multicolumn{2}{c}{attainable saving} \\
\cmidrule(lr){3-4}
Region profile & $\ell_r$ range & common floor & solved for $\ell_r$ \\
\midrule
Mild, $6/6/6/3/3$ & $0.875$--$0.895$ & $37.5\%$ & $36.9\%$ \\
Severe, $15/5/2/1/1$ & $0.777$--$0.908$ & $79.2\%$ & $78.2\%$ \\
Consortium by country & $0$--$0.899$ & $88.0\%$ & $83.2\%$ \\
\bottomrule
\end{tabular}
\end{table}

\subsection{Exposure in operational deployments}

Since $\delta$ is a property of a deployment rather than of an experiment, it can
be measured directly for real federations, without reference to any model or
dataset. Table~\ref{tab:realdelta} states each structure completely, since
$\delta$ depends on the silo weights as well as the region sizes, and the two must
be given together.

One special case is worth recording, because it explains the resulting
magnitudes. When all silos carry equal weight, $\rho_r = 1/m_r$ and
\eqref{eq:delta} reduces to
\begin{equation}
\delta \;=\; 1 - \frac{|\Reg| \, m_{\min}}{n},
\label{eq:deltaequal}
\end{equation}
with $m_{\min}$ the smallest region. The saving is then governed by the presence
of one small region alongside a large federation, and it approaches one as $n$
grows with $|\Reg|$ fixed. The consortium and cellular entries are equal-weight
structures and follow \eqref{eq:deltaequal} directly; the clinical entry uses the
measured site sizes of a public benchmark, which is why it is far lower than its
region profile alone would suggest. Federations whose regions are equally sized
give exactly zero, as do flat federations with no aggregation tier. These values
locate a deployment on the same $\delta$ axis along which Figures~\ref{fig:gain}
and~\ref{fig:gainlog} report the resulting gain.

\begin{table}[t]
\caption{Exposure dispersion for operational structures, with the weights each is computed from. Equal-weight rows follow \eqref{eq:deltaequal}; the clinical row uses measured site sizes, and its $\delta$ would be $33.3\%$ were those sites equally sized.}
\label{tab:realdelta}
\centering
\footnotesize
\setlength{\tabcolsep}{4pt}
\begin{tabular}{@{}p{0.30\columnwidth}p{0.24\columnwidth}p{0.20\columnwidth}c@{}}
\toprule
Structure & Regions & Silo weights & $\delta$ \\
\midrule
Consortium by country & $30/10/5/3/1/1$ & equal & $88.0\%$ \\
Cellular edge, 60 base stations \cite{wang2017cellular} & lognormal, $\mu{=}1.2$, $s{=}1.3$, floor 1 & equal & $82.8$--$87.6\%$ \\
Clinical sites by hospital (Fed-ISIC2019 \cite{terrail2022flamby}) & $3/1/1/1$ & $9930$, $3163$, $2691$, $1807$, $655$, $351$ & $14.4\%$ \\
Equally sized regions & any, equal & any & $0\%$ \\
Flat federation & one silo each & any & $0\%$ \\
\bottomrule
\end{tabular}
\end{table}

These figures follow directly from how each deployment is organised. Cellular
deployments sit high on the scale because base station loads are long tailed, with
dense cells serving many devices and sparse cells very few. Consortia organised by
country behave similarly, since site counts per country are uneven. Having
established what the theory predicts for structures like these, we turn next to
testing that prediction directly.

\section{Experimental Evaluation}
\label{sec:eval}

\subsection{Design}

Corollary~\ref{cor:delta} makes a prediction that experiment can contradict, and
we designed the evaluation around that possibility rather than around a favourable
configuration. We sweep the exposure profile across the range of $\delta$,
including a balanced deployment at $\delta = 0$ for which the theory predicts
parity with the baseline. A second balanced configuration, at four regions rather
than six, was run throughout and reproduced the same values; we report only one of
the two, since under equal weights both are identically zero. Three predictions
are stated in advance.

The first prediction is that the budget saved at a matched privacy target equals
$\delta$. The second is that the accuracy gain at matched privacy increases with
$\delta$ and reaches zero when $\delta$ does. The third is that the gain depends
on allocating according to exposure rather than on being non-uniform at all, which
we test in two ways: against an allocation that preserves the dispersion of noise
while assigning it to the wrong regions, and against size-based heuristics of the
kind a practitioner might reach for without the analysis.

Every arm holds each silo at the same worst-case bound $K$, so the comparison
fixes the guarantee received by the least protected silo, and the arms differ only
in the total noise they require to reach it. Silo $i$ is concealed by the total
noise in its region, so its effective noise multiplier is $\sqrt{S_r}/w_i$, from
which $\varepsilon$ follows.

What the worst case fixes, it fixes only at the extreme, and the distribution
behind it still moves. Uniform allocation reaches $K$ by giving every silo the
noise the most exposed region requires, which leaves silos in well concealed
regions strictly inside the target; the optimal allocation brings each region to
$K$ exactly. Table~\ref{tab:epsdist} reports the consequence for the severe
profile. Holding the maximum at $\varepsilon = 0.99$, the mean rises from $0.368$
to $0.990$, and the largest region moves from $0.229$ to $0.990$. The budget the
allocation frees is therefore drawn from silos that previously held protection
beyond the stated target, and equalisation is what releases it. Read the other
way, the uniform multiplier delivers guarantees spanning a factor of $4.3$ across
silos, whereas the optimal allocation delivers the target to all of them.

\begin{table}[t]
\caption{Per-silo $\varepsilon$ by region under each allocation, severe profile $15/5/2/1/1$ at $n = 96$, $T = 10$, $\delta_{\mathrm{DP}} = 10^{-5}$. The worst case is common to both arms by construction; the distribution behind it is not.}
\label{tab:epsdist}
\centering
\footnotesize
\setlength{\tabcolsep}{4pt}
\begin{tabular}{@{}lcccccc@{}}
\toprule
& \multicolumn{5}{c}{$\varepsilon$ by region (silos per region)} & \\
\cmidrule(lr){2-6}
Allocation & $60$ & $20$ & $8$ & $4$ & $4$ & mean \\
\midrule
Uniform            & $0.229$ & $0.414$ & $0.680$ & $0.990$ & $0.990$ & $0.368$ \\
Optimal (\Fulcrum) & $0.990$ & $0.990$ & $0.990$ & $0.990$ & $0.990$ & $0.990$ \\
\bottomrule
\end{tabular}
\end{table}

\subsection{Datasets and models}

We evaluate on two modalities to establish that the conversion from budget to
accuracy generalises across data types.

For images we use CIFAR-10 \cite{krizhevsky2009cifar}, grouped into two
superclasses so that the sensitive-class concentration is a binary proportion. We
embed images with a ResNet-18 \cite{he2016resnet} pretrained on ImageNet
\cite{deng2009imagenet}, holding the backbone frozen, and reduce the resulting
representation to $32$ dimensions by principal component analysis fitted on the
training split.

For text we use AG News \cite{zhang2015agnews}, likewise reduced to two classes.
We embed documents with the all-MiniLM-L6-v2 sentence encoder
\cite{reimers2019sentencebert,wang2020minilm}, again frozen, and apply the same
dimensionality reduction.

In both cases a linear classification head is trained federatively on the frozen
representation, giving an identical head dimension across the two tasks. Holding
this dimension fixed keeps the signal-to-noise regime comparable and makes the
two sets of results directly interpretable against one another.
Section~\ref{sec:discussion} explains why fine-tuning a pretrained representation,
rather than training a network from scratch, is the appropriate setting for
silo-level privacy.

\subsection{Configuration}

With the representations fixed, the remaining experimental parameters are as
follows. We use $n = 96$ silos partitioned into regions according to the profile
under test, $T = 10$ rounds, $\delta_{\mathrm{DP}} = 10^{-5}$, clipping at $C = 1$,
ten local optimisation steps per round, a batch size of $64$, and ten independent
seeds per configuration. Local class distributions are generated by a coupled
partitioning scheme in which a region-determined component is mixed with a random
component, so that concentrations vary across silos in a controlled manner while
dataset sizes remain equal by construction.

\begin{figure}[t]
\centering
\includegraphics{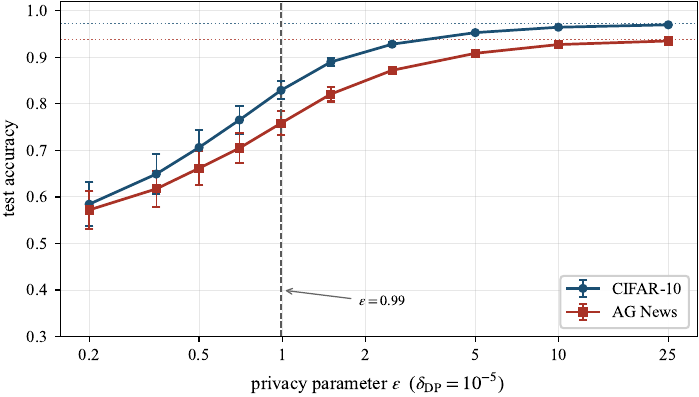}
\caption{Accuracy against the per-client privacy parameter for both modalities
under uniform allocation on the balanced six-region deployment, with non-private
references shown dotted. Error bars give the standard error over ten seeds. The
evaluation operates at $\varepsilon = 0.99$, where accuracy retains sensitivity to
the noise level and the comparison between allocations is therefore meaningful.}
\label{fig:pu}
\end{figure}

The privacy target is $K = 0.88$ nats, and it yields $\varepsilon = 0.99$ in every
run reported here. Figure~\ref{fig:pu} shows the privacy-utility relationship
across a wider range and marks the operating point.

Remark~\ref{rem:informative} requires $T \cdot m_i(\sigma) < H(p_i) - \ell_i$ for
the guarantee to convey anything, and we verify it here. Over the $3840$ silo
instances per task across all profiles and seeds, the concentrations $p_i$ span
$[0.173, 0.827]$, giving $H(p_i) = 2.23$ nats at $R = 20$ and a margin $H(p_i) -
\ell_i = 1.38$ nats. The mechanism term reaches $K - \ell_i = 0.030$ nats at its
largest, which is $2.2\%$ of that margin, so the condition holds with two orders
of magnitude to spare at every operating point and under every allocation.

Because $H(p_i)$ grows with $R$, the test would pass trivially at fine resolution,
so Table~\ref{tab:informative} reports it across $R$ for CIFAR-10, with AG News
agreeing to within $0.02$ nats at every resolution. The condition holds at every
resolution down to $R = 4$, where the target is resolved only to quartiles and
$H(p_i) = 1.33$ nats still exceeds $K$. The verification therefore does not depend
on the choice of $R$.

\begin{table}[t]
\caption{Informativeness of the guarantee across quantisation resolutions. The condition of Remark~\ref{rem:informative} requires the mechanism term, at most $0.030$ nats here, to fall below the margin $H(p_i) - \ell_i$. It holds at every resolution, including the coarsest.}
\label{tab:informative}
\centering
\footnotesize
\setlength{\tabcolsep}{5pt}
\begin{tabular}{@{}cccccc@{}}
\toprule
$R$ & resolution & $H(p_i)$ & margin & mech.\ term & informative \\
\midrule
$4$   & $0.250$ & $1.33$ & $0.48$ & $0.030$ & yes \\
$10$  & $0.100$ & $1.64$ & $0.79$ & $0.030$ & yes \\
$20$  & $0.050$ & $2.23$ & $1.38$ & $0.030$ & yes \\
$50$  & $0.020$ & $3.04$ & $2.19$ & $0.030$ & yes \\
\bottomrule
\end{tabular}
\end{table}

\subsection{Results}

\begin{figure}[t]
\centering
\includegraphics{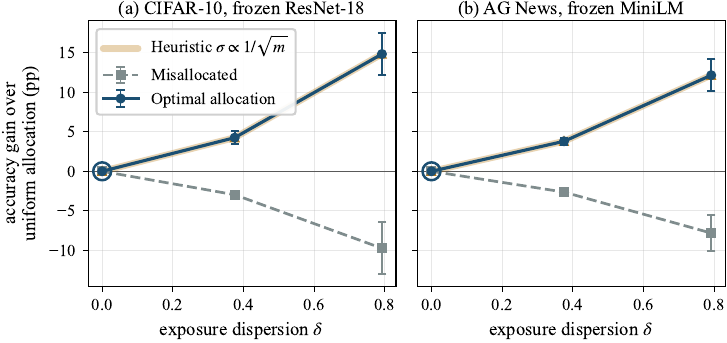}
\caption{Equally sized silos. Accuracy gain over uniform allocation at a matched
worst-case per-client guarantee, against exposure dispersion. Here $\rho_r = 1/m_r$,
so the size heuristic and the optimal allocation are the same rule and agree to
machine precision on every seed; the heuristic is drawn as a wide translucent band
because it would otherwise be hidden beneath the optimum. The circled marker is the balanced
control, for which the theory predicts parity and which delivers exactly zero. Because the arms are paired within each seed, error bars give the standard
error of the paired difference over ten seeds.}
\label{fig:gain}
\end{figure}

\begin{figure}[t]
\centering
\includegraphics{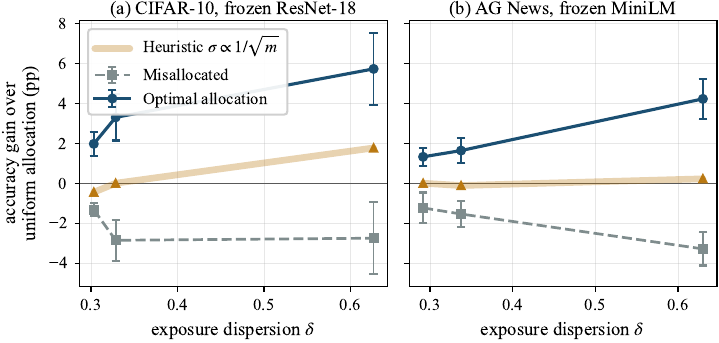}
\caption{Log-normal silo sizes, spanning roughly an order of magnitude. The
identity $\rho_r = 1/m_r$ no longer holds, and the optimal allocation separates
from the size heuristic. Exposure dispersion is recomputed from the realised
weights, so the same region profiles sit at different $\delta$ than in
Figure~\ref{fig:gain}: the balanced profile rises above zero, because equally sized
regions still differ in $\rho_r$ once weights differ, while the severe profile
falls, because a dominant silo pushes its region's $\rho_r$ toward one and narrows
the spread. Error bars as in Figure~\ref{fig:gain}.}
\label{fig:gainlog}
\end{figure}

Table~\ref{tab:main} reports the outcome, which Figure~\ref{fig:gain} plots for
equally sized silos and Figure~\ref{fig:gainlog} for log-normal ones.

Budget savings match $\delta$ to three decimal places in every configuration,
confirming the first prediction and Corollary~\ref{cor:delta}.

Matching budgets is a necessary check on the theory, but what those budgets buy in
accuracy is the second and more consequential prediction. The accuracy gain
increases with $\delta$, reaching $14.84 \pm 2.68$ points on images and $12.16 \pm
2.00$ points on text in the most heterogeneous deployment, quoted as means over
ten seeds with the standard error of the paired difference. In the balanced
deployment the gain equals zero to the precision of the measurement on both
tasks, confirming the second prediction in full. This is the outcome the theory
requires, and we regard it as the most informative row in the table, since a
method acting on some quantity other than the one we claim would be unlikely to
produce exactly zero here.

The third prediction is likewise confirmed. Given the same dispersion of noise but
assigned to the wrong regions, and rescaled to meet the same target, the
misallocation control performs below uniform allocation, by roughly three points
in the mild configuration and by eight to ten points in the severe one.
Non-uniformity alone is therefore insufficient, and the gain depends on allocating
according to the exposure profile specifically.

Beyond confirming the three predictions, the results show a further regularity
across tasks. Agreement between the two modalities is close, at $4.25$ against
$3.80$ points in the mild configuration and $14.84$ against $12.16$ points in the
severe one. We had anticipated that the magnitudes would differ, since the gain
depends on the local slope of accuracy against noise, which is a property of the
task. The close agreement is most plausibly a consequence of having matched the
head dimension and privacy target across the two tasks, and we would expect
divergence were either varied. Having confirmed the prediction experimentally, we
turn next to the assumptions and scope behind it.

\begin{table}[t]
\caption{Accuracy at a matched worst-case per-client guarantee ($\varepsilon = 0.99$, $n = 96$, $T = 10$, ten seeds). Gains are percentage points relative to uniform allocation, and the budget saved equals $\delta$ in every row. Non-private references are $0.972$ for CIFAR-10 and $0.938$ for AG News.}
\label{tab:main}
\centering
\footnotesize
\setlength{\tabcolsep}{3pt}
\begin{tabular}{@{}lc cccc cccc@{}}
\toprule
& & \multicolumn{4}{c}{CIFAR-10 (ResNet-18)} & \multicolumn{4}{c}{AG News (MiniLM)} \\
\cmidrule(lr){3-6}\cmidrule(lr){7-10}
Region profile & $\delta$
& \Fulcrum & Unif. & Gain & Misall.
& \Fulcrum & Unif. & Gain & Misall. \\
\midrule
Balanced, 6 regions & $0.000$
& $0.829$ & $0.829$ & $+0.00$ & $+0.00$
& $0.758$ & $0.758$ & $+0.00$ & $+0.00$ \\
Mild, $6/6/6/3/3$ & $0.375$
& $0.841$ & $0.798$ & $+4.25$ & $-2.97$
& $0.773$ & $0.735$ & $+3.80$ & $-2.63$ \\
Severe, $15/5/2/1/1$ & $0.792$
& $0.843$ & $0.695$ & $\mathbf{+14.84}$ & $-9.75$
& $0.774$ & $0.652$ & $\mathbf{+12.16}$ & $-7.82$ \\
\bottomrule
\end{tabular}
\end{table}

\subsection{Comparison against size-based heuristics}

Uniform allocation is the incumbent, but it is not the most informed comparator. A
practitioner who understands that a silo is concealed by its region, without
deriving how much, would scale noise by region size. Two exponents are natural.
Trusted-aggregator schemes inject one noise draw on behalf of a zone and therefore
require $\sigma \propto 1/m$; ported directly to local injection, this gives the
first heuristic. The second, $\sigma \propto 1/\sqrt{m}$, is the exponent implied
by Remark~\ref{rem:regional} when each silo injects its own noise. We rescale both
to meet the same target $K$ and evaluate them alongside the allocation.

One identity governs the comparison. When all silos carry equal aggregation
weight, $\rho_r = W_r/V_r = 1/m_r$ exactly, so \eqref{eq:alloc} reduces to
$\sigma_i \propto 1/\sqrt{m_r}$. The second heuristic is therefore not a rival to
the allocation but its equal-weight special case, and the upper half of
Table~\ref{tab:baselines} shows the two agreeing to four decimal places, at
$+4.25$ and $+14.84$ points. Equality of weights is a knife-edge condition, and
the identity fails as soon as silo sizes differ at all.

They differ in every federation we are aware of. The lower half of
Table~\ref{tab:baselines} draws silo sizes log-normally, giving roughly an order of
magnitude between largest and smallest. The balanced profile isolates what the
allocation adds: its regions are equally sized, so a size-based rule is by
construction indistinguishable from uniform and returns exactly zero, while the
allocation gains $3.29$ points on images and $1.32$ on text by responding to
weight concentration that region size alone cannot express. On the severe profile
it exceeds the heuristic by $3.96$ points on images and $4.00$ on text. Across
every profile and both modalities the allocation is at least as accurate as the
heuristic and strictly more accurate wherever silo sizes vary, which is exactly
what \eqref{eq:alloc} being the optimum of \eqref{eq:opt} requires. No regime
therefore favours the heuristic: it costs the same to compute and forfeits the
gain precisely when the federation is heterogeneous.

The naive $1/m$ port behaves worse than uniform in most heterogeneous
configurations, reaching $-9.39$ points. This is the practical consequence of the
distinction drawn in Section~\ref{sec:related}: an exponent calibrated for a
trusted aggregator injecting once per zone is not the exponent for silos injecting
independently, and transferring it costs more than making no adjustment at all.

Silo weights vary in practice, and the extent of that variation is well
documented. Unbalanced client dataset sizes are treated as a defining feature of
the federated setting rather than an incidental one \cite{kairouz2021advances},
and the seven healthcare federations assembled in FLamby
\cite{terrail2022flamby} bear this out without exception. Per-centre sample
counts there differ by factors between $1.6$ and $28.3$, with log standard
deviations from $0.23$ to $1.09$ and a median of $0.59$. The log-normal arm
evaluated above draws silo sizes at a log standard deviation of $0.60$, which
places it at the median of the seven rather than at a favourable extreme.

Table~\ref{tab:realdelta} nonetheless reports the consortium and cellular
structures with equal weights, since their sources describe how many sites belong
to each region without recording how much data each site holds, whereas the
Fed-ISIC2019 entry uses the published per-centre volumes. What that dispersion
costs a size-based rule depends on how many silos share a region. Fed-ISIC2019
places at most three silos in a region, and meeting the same guarantee there
through $\sigma \propto 1/\sqrt{m}$ costs $1.06$ times the optimal budget, a
margin of no practical interest. Applying the median and the most skewed of the
seven dispersions to the consortium and cellular structures, which place tens and
hundreds of silos in a region, costs $2.20$ to $2.65$ times and $3.71$ to $4.50$
times the optimal budget, while $\delta$ falls from $88.0\%$ to between $44.0\%$
and $66.0\%$ and from $85.6\%$ to between $34.4\%$ and $61.0\%$. Equal weights
maximise $\delta$ and are the case in which the heuristic attains it, whereas
dispersed weights lower $\delta$ and leave the allocation as the only rule
reaching what remains. Since the allocation costs no more to compute and is never
less accurate, what varies is not which rule to prefer but how much the choice
matters, and that grows with the size of the federation.

\begin{table}[t]
\caption{Accuracy gain over uniform allocation (percentage points, paired over ten seeds) for the allocation and three alternatives, at a matched worst-case per-client guarantee. Upper half: equally sized silos, where $\rho_r = 1/m_r$ and the $1/\sqrt{m}$ heuristic coincides with the optimum. Lower half: log-normal silo sizes, where the two separate. The balanced profiles in the lower half have equally sized regions, so any size-based rule is identical to uniform there.}
\label{tab:baselines}
\centering
\footnotesize
\setlength{\tabcolsep}{2.6pt}
\begin{tabular}{@{}llc cccc cccc@{}}
\toprule
& & & \multicolumn{4}{c}{CIFAR-10 (ResNet-18)} & \multicolumn{4}{c}{AG News (MiniLM)} \\
\cmidrule(lr){4-7}\cmidrule(lr){8-11}
Silo sizes & Region profile & $\delta$
& \Fulcrum & $1/\sqrt{m}$ & $1/m$ & Misall.
& \Fulcrum & $1/\sqrt{m}$ & $1/m$ & Misall. \\
\midrule
\multirow{3}{*}{Equal}
 & Balanced, 6 regions & $0.000$ & $+0.00$ & $+0.00$ & $+0.00$ & $+0.00$ & $+0.00$ & $+0.00$ & $+0.00$ & $+0.00$ \\
 & Mild, $6/6/6/3/3$   & $0.375$ & $+4.25$ & $+4.25$ & $+1.24$ & $-2.97$ & $+3.80$ & $+3.80$ & $+0.94$ & $-2.63$ \\
 & Severe, $15/5/2/1/1$& $0.792$ & $\mathbf{+14.84}$ & $\mathbf{+14.84}$ & $-3.15$ & $-9.75$ & $\mathbf{+12.16}$ & $\mathbf{+12.16}$ & $-2.79$ & $-7.82$ \\
\midrule
\multirow{3}{*}{Log-normal}
 & Balanced, 6 regions & $\sim\!0.31$ & $\mathbf{+3.29}$ & $+0.00$ & $+0.00$ & $-2.86$ & $\mathbf{+1.32}$ & $+0.00$ & $+0.00$ & $-1.23$ \\
 & Mild, $6/6/6/3/3$   & $\sim\!0.32$ & $\mathbf{+1.97}$ & $-0.45$ & $-2.34$ & $-1.33$ & $\mathbf{+1.64}$ & $-0.12$ & $-1.90$ & $-1.54$ \\
 & Severe, $15/5/2/1/1$& $\sim\!0.63$ & $\mathbf{+5.72}$ & $+1.76$ & $-8.89$ & $-2.75$ & $\mathbf{+4.22}$ & $+0.22$ & $-9.39$ & $-3.28$ \\
\bottomrule
\end{tabular}
\end{table}

\section{Discussion}
\label{sec:discussion}

The experiments just reported were run in a specific setting, namely fine-tuning
a frozen representation, and that choice is deliberate. We explain why it is
appropriate before turning to guidance, scope, and limitations.

\subsection{Why fine-tuning is the appropriate setting}
\label{sec:finetune}

The noise entering the aggregate has norm proportional to $C \sigma \sqrt{d
\sum_i w_i^2}$, where $d$ counts trainable parameters, while the aggregate signal
has norm proportional to $C \sqrt{a}$ for an average alignment $a$ among the
clipped updates. Their ratio therefore behaves as $\sqrt{a n / d} \, / \, \sigma$.
Parameter count enters directly, and at silo-level granularity it governs
feasibility.

The consequence is sharp. A convolutional network trained from scratch in our
setting carries $d$ on the order of $5 \times 10^5$ against $n = 96$ silos, and at
noise levels for which the bound remains informative the model stays at chance
accuracy. Fine-tuning a compact head on a frozen backbone reduces $d$ by roughly
four orders of magnitude and recovers most of the non-private accuracy, reaching
$85\%$ of it on images and $81\%$ on text at the operating point and converging to
it by $\varepsilon \approx 10$, as Figure~\ref{fig:pu} shows. We therefore present
silo-level differential privacy as suited to federated fine-tuning of pretrained
representations, which is also where much current practice already sits. The
allocation result itself holds irrespective of this choice, since it concerns
relative exposure rather than absolute feasibility.

\subsection{Behaviour at higher model dimension}
\label{sec:dimension}

The setting evaluated here trains a compact head, whereas production systems often
tune more parameters, whether fully or through adapters. Two quantities respond
differently to that change, and separating them answers what happens.

The dispersion $\delta$ is unaffected. Equation~\eqref{eq:delta} is a function of
the aggregation weights and the region structure alone, with no dependence on the
model, the data, or the number of parameters, so the budget recovered at a fixed
guarantee is identical at any $d$. What responds is the accuracy that budget buys,
through the signal-to-noise ratio $\sqrt{an/d}\,/\,\sigma$ of
Section~\ref{sec:finetune}: raising $d$ moves every arm down its own
accuracy-versus-noise curve, compressing the differences between them.

Repeating the CIFAR-10 sweep with a head of $d = 258$ against the $d = 66$ used
elsewhere bears this out. The budget saved is unchanged to the digit, at
$37.500\%$ and $79.167\%$ on the mild and severe profiles in both cases, while the
accuracy gain falls from $4.25$ to $4.04$ points and from $14.84$ to $13.51$
points as the uniform baseline itself declines from $0.798$ to $0.759$ and from
$0.695$ to $0.663$. The privacy claim therefore transfers to larger heads
unchanged, and the utility realised from it shrinks as the model moves toward the
noise floor. A deployment tuning many parameters at silo-level granularity is
constrained by that floor rather than by the allocation, the same conclusion
Section~\ref{sec:finetune} draws for training from scratch.

\subsection{Practical guidance}

Beyond justifying the experimental setting, the same result yields direct
operational guidance. The workflow it suggests is brief. A practitioner computes
$\rho_r$ for each region from the deployment description, forms $\delta$, and
reads off the budget that reallocation recovers. Where $\delta$ is small, the
uniform multiplier already sits close to optimal and little is available to
recover. Where $\delta$ is large, \Fulcrum (Algorithm~\ref{alg:main}) yields the
allocation at negligible computational cost, since $K^\star$ is obtained by
one-dimensional bisection performed once before training.

Identifying the regime in advance is unnecessary, because the allocation recovers
the simpler rules rather than competing with them. It returns the uniform
multiplier when regions are equally exposed, and the $1/\sqrt{m}$ rule of
Section~\ref{sec:eval} when silos carry equal weight. Running it is therefore
never worse than running either, and it is better whenever the deployment departs
from those conditions.

The deployment measurements of Section~\ref{sec:allocation} suggest that many
operational deployments fall into the second category. A federation seeking to
reduce its exposure to this effect altogether has a complementary option, which is
to balance its regions when the structure is under its control, since equally
sized regions give $\delta = 0$ and leave nothing for reallocation to recover.

\subsection{Composition with secure aggregation}
\label{sec:secagg}

Section~\ref{sec:adversary} assumes an observer positioned above the regional
tier, and Section~\ref{sec:limitations} identifies the resulting trust placed in
regional aggregators as the sharpest limitation of the scheme. Secure aggregation
\cite{bonawitz2017secagg} is the standard means of removing it, and the
allocation composes with it without modification.

Where the masking sits determines whether anything changes, and it is worth
separating two placements that are easily conflated.

Masking \emph{within} a region, so that only $Y_r^{(t)}$ is recoverable from its
members, maps onto the framework exactly. Theorem~\ref{thm:bound} already
conditions on the view $\{Y_r^{(t)}\}$, and this form of masking changes which
parties are restricted to that view rather than changing the view itself. Every
input to the optimisation is therefore untouched: the per-region weight statistics
$W_r$ and $V_r$ are computed from the same membership, the mechanism term
\eqref{eq:mterm} keeps the same denominator, and the allocation \eqref{eq:alloc},
the budget equation and the dispersion \eqref{eq:delta} are unchanged. Numerically
nothing moves, because the masks cancel within each region and the aggregate is
identical with and without them. What the deployment gains is that the regional
aggregator is now held to the observer model as well, so the per-silo guarantee
stops degrading against it, the very liability identified in
Section~\ref{sec:adversary}.

Masking \emph{across} regions, so that the central aggregator recovers only the
global sum $\sum_r Y_r^{(t)}$, is a different proposition and does change the
inputs. The concealment set becomes the entire federation rather than one region,
so the mechanism term reduces to $m_i(\sigma) = 2w_i^2 / \sum_{j} w_j^2 \sigma_j^2$
with the denominator running over all $n$ silos. The bound then depends on the
noise only through the global total, every allocation carrying the same total is
optimal, and $\delta = 0$ by the same argument that gives $\delta = 0$ for a flat
federation. The federation behaves as a single region.

Taken together these two cases delimit where the allocation is informative. It
requires an observer that resolves per-region sums and no finer: a flat
federation exposes each silo individually and leaves nothing to pool, while global
masking pools everything and leaves nothing to distribute. The intended
composition is therefore within-region masking, which retains the structure the
allocation acts on while removing the trust that structure would otherwise demand.

Two qualifications follow from how masking behaves in practice. A protocol
completing only when at least $t$ of $m$ silos report will drop rounds below that
threshold; dropped rounds release nothing and so reduce the number of mechanism
invocations, leaving the guarantee for completed rounds intact. More consequential
is collusion. If a set $\mathcal{C} \subset r$ of silos shares its own masks and
noise with the observer, their contributions can be subtracted from $Y_r^{(t)}$,
and silo $i$ is then concealed only by the honest remainder, so \eqref{eq:mterm}
holds with $V_r$ replaced by
\begin{equation}
V_r^{\mathrm{hon}} \;=\; \textstyle\sum_{j \in r \setminus \mathcal{C}} w_j^2 .
\end{equation}
The allocation is structurally identical under this substitution; only the
definition of the exposure ratio changes, to
$\rho_r = W_r / V_r^{\mathrm{hon}}$. A deployment provisioning against a coalition
of up to $c$ silos per region therefore computes $\delta$ from the honest
remainder and obtains a smaller but still explicit saving. In the limiting case
where all but one member of a region may collude, $\rho_r \to 1$ for every region
and $\delta \to 0$, which is the flat-federation regime of
Section~\ref{sec:allocation}.

\subsection{Sensitivity to the assumed coupling}

Just as secure aggregation removes the trust assumption on the observer, a
further assumption remains in the generative model behind the lateral floor. A
deployment whose correlation structure departs from it realises a different
floor. The effect on the achieved guarantee is exact and does not compound.
Allocating with $\ell_r(\kappa_{\mathrm{a}})$ while the deployment realises
$\ell_r(\kappa_{\mathrm{t}})$ gives $S_r = aW_r/(K - \ell_r(\kappa_{\mathrm{a}}))$,
so the bound actually attained is
\begin{equation}
\frac{aW_r}{S_r} + \ell_r(\kappa_{\mathrm{t}})
\;=\; K + \big[\ell_r(\kappa_{\mathrm{t}}) - \ell_r(\kappa_{\mathrm{a}})\big].
\label{eq:missp}
\end{equation}
The bracket in \eqref{eq:missp} is the whole of the effect: the target is
overshot by exactly the error in the floor, entering additively, and the
allocation neither amplifies nor attenuates it.

Three consequences are worth stating. Overestimating the coupling is free, since
$\ell_r(\kappa_{\mathrm{t}}) \le \ell_r(\kappa_{\mathrm{a}})$ makes the bracket
non-positive and the target is still met. Underestimating it costs the gap and no
more: assuming $\kappa = 5$ where the deployment realises $\kappa = 100$
overshoots $K = 2.2$ nats by $1.13$ nats on the severe profile, and assuming
$\kappa = 20$ under the same conditions overshoots by $0.70$. The overshoot is
bounded in all cases by $\max_\kappa I(p_i;\Phi_g)$, which is $1.62$ nats for the
family considered here, because the floor cannot exceed its own saturation
ceiling. A practitioner uncertain about the correlation structure should
therefore assume the strongest coupling considered plausible, which costs nothing
when it turns out to be an overestimate and bounds the exposure when it is not.

\subsection{Limitations}
\label{sec:limitations}

Several assumptions are stated where they are used: disjoint regions with fixed
membership and disjoint silo datasets (Section~\ref{sec:model}), the trust the
concealment places in regional aggregators (Sections~\ref{sec:adversary}
and~\ref{sec:secagg}), and the requirement that regions differ in effective size,
which $\delta$ tests in advance (Section~\ref{sec:allocation}). Three further
boundaries delimit the result.

The first is that the analysis is conservative. Theorem~\ref{thm:bound} is an
upper bound carrying slack in at least two places, since the chain rule discards a
non-negative term and the per-round divergence is a worst case over adjacent
inputs. A tighter analysis would meet the same target with less noise, so the
accuracy reported in Section~\ref{sec:eval} understates what the mechanism can
actually deliver. The allocation itself is indifferent to this, since it depends
on the relative exposure of regions rather than on the absolute level of the
bound, so any sharpening of the analysis transfers to the mechanism unchanged.

The second is the position of the adversary, which is passive: it follows the
protocol and reads only what its position exposes. An aggregator that misreports
the sum it computed, and active manipulation of the protocol more generally, lie
outside this model; poisoning conducted under the cover of the noise the
mechanism introduces \cite{zheng2025dppoison} is a concrete instance. Local
perturbation already removes the need to trust the aggregator to add noise, and
Section~\ref{sec:secagg} sets out how secure aggregation removes the remaining
need to trust it to report honestly, together with what that costs under
collusion.

The third is the reach of the evidence presented. We evaluate two modalities at
one federation scale, with a linear head on frozen representations over ten
rounds. That regime carries further than its size suggests, since
Section~\ref{sec:dimension} establishes that the budget recovered is invariant to
model dimension while the accuracy it buys compresses, which fixes the ordering of
the arms at larger heads as well. Larger silo counts, longer training, and dynamic
membership remain to be characterised. One property bears stating precisely:
composition in the information bound is linear in $T$, which is looser than the
R\'enyi accounting behind the reported $\varepsilon$, so the bound degrades faster
with round count than the privacy budget itself does.

\subsection{Relation to trusted-aggregator schemes}

The trust assumed of the aggregator, touched on above, is worth situating in more
detail. As noted in Section~\ref{sec:related}, a trusted super-node that
contributes a single noise draw on behalf of its region requires per-client noise
scaling as $1/m$ in place of $1/\sqrt{m}$. A deployment able to trust its
aggregators therefore reaches the same guarantee with less noise, and the factor
$\sqrt{m}$ measures what the untrusted setting costs. The allocation problem
arises in both regimes, and Theorem~\ref{thm:allocation} applies to either once
the mechanism term is written accordingly, while the value of $\delta$ in
Corollary~\ref{cor:delta} is derived specifically for the untrusted case treated
here. Having situated the result against its assumptions and its alternatives, we
close by drawing the argument together.

\section{Conclusions and Future Work}
\label{sec:conclusion}

In hierarchical federated learning, a participant is concealed by the region it
is aggregated with, and regions in operational deployments differ substantially
in effective size. A single shared noise multiplier must be calibrated for the
most exposed region and therefore leaves the remainder carrying surplus noise. We
gave a silo-level differential privacy guarantee for the mechanism and derived a
separate per-client information bound whose adjacency matches the property being
protected. Solving the resulting allocation problem explicitly yields \Fulcrum,
whose saving relative to uniform allocation, where the lateral floors agree, is
exactly $\delta = 1 - \langle \rho \rangle_V / \rho_{\max}$, a quantity computable
from the deployment in advance of training and equal to zero precisely when all
regions are equally exposed. Measured across operational structures, $\delta$
reaches $88\%$. Experiments on image and text classification confirmed the
prediction at $\varepsilon = 0.99$, including a control deployment for which the
theory predicts parity and which delivered it.

Three directions follow naturally. An allocation that combines structural
exposure with declared per-participant preferences would unify this work with
personalised differential privacy. A further step would couple it with
personalisation of the model itself, which decentralised federated learning has
approached through per-node uncertainty estimates
\cite{rangwala2026evidential}. Extending the analysis to overlapping regions
would cover deployments in which a silo contributes to more than one aggregate.
Combining the allocation with secure aggregation would relax the remaining trust
assumption on the aggregator while preserving the structure of the result.

\section*{Data and Code Availability}

The \Fulcrum source code, together with the analysis artifacts underlying every
table and figure is available at \url{https://doi.org/10.5281/zenodo.20507155}.

\begin{acks}
The authors thank Prof. Salil Kanhere at the University of New South Wales for his valuable feedback
during the preparation of this work. This work is supported by the Australian Research Council (ARC) through Discovery Project grant DP240102088.
\end{acks}

\bibliographystyle{ACM-Reference-Format}
\bibliography{references}

\appendix

\section{Proof of Theorem~\ref{thm:bound}}
\label{app:bound}

The argument proceeds in four movements. We first reduce the adversary's estimate
to the observations it was formed from, so that nothing depends on how clever the
adversary is. We then split the observations into the part explained by the other
silos' data, which is the lateral floor, and the part that remains, which is what
the mechanism must control. We next isolate a single silo's contribution within a
single round, which is where conditioning on the past does the essential work.
Finally, the per-round contribution is bounded by the largest divergence the
Gaussian mechanism can produce between two silo-adjacent inputs, and these bounds
accumulate across rounds.

Two standard facts are used repeatedly \cite{cover2006elements}. The first lets us
trade a marginal dependence for a conditional one, and the second converts a
worst-case divergence into a bound on mutual information.

\begin{lemma}[Chain rule bound]
\label{lem:chain}
For random variables $X, Y, Z$, $I(X;Y) \le I(X;Z) + I(X;Y \mid Z)$.
\end{lemma}
\begin{proof}
Expanding $I(X;Y,Z)$ in two ways gives $I(X;Z) + I(X;Y \mid Z) = I(X;Y) + I(X;Z
\mid Y)$, and $I(X;Z \mid Y) \ge 0$.
\end{proof}

\begin{lemma}[Maximal divergence bound]
\label{lem:maxkl}
If a channel $P_{Y \mid X}$ satisfies $\KL(P_{Y \mid x} \,\|\, P_{Y \mid x'}) \le
\kappa$ for all $x, x'$, then $I(X;Y) \le \kappa$.
\end{lemma}
\begin{proof}
$I(X;Y) = \mathbb{E}_X \KL(P_{Y \mid X} \| P_Y)$ with $P_Y = \mathbb{E}_{X'} P_{Y
\mid X'}$. By convexity of the divergence in its second argument, $\KL(P_{Y \mid
x} \| P_Y) \le \mathbb{E}_{X'} \KL(P_{Y \mid x} \| P_{Y \mid X'}) \le \kappa$.
\end{proof}

\begin{proof}[Proof of Theorem~\ref{thm:bound}]
\emph{Step 1: reduce the estimate to the observations.}
Since $\hat{p}_i$ is a deterministic function of $Y$, the data processing
inequality gives $I(p_i ; \hat{p}_i) \le I(p_i ; Y)$. Every subsequent step
therefore bounds a quantity that no choice of adversary strategy can exceed.

\emph{Step 2: separate the lateral floor.} Applying
Lemma~\ref{lem:chain} with $Z = D_{-i}$,
\begin{equation}
I(p_i ; Y) \le I(p_i ; D_{-i}) + I(p_i ; Y \mid D_{-i})
= \ell_i + I(p_i ; Y \mid D_{-i}).
\end{equation}
Because $p_i$ is a function of $D_i$, a further application of the data processing
inequality gives $I(p_i ; Y \mid D_{-i}) \le I(D_i ; Y \mid D_{-i})$.

The first term here is the lateral floor, which the mechanism leaves untouched.
The second is the quantity the noise must control, and the remainder of the proof
bounds it.

\emph{Step 3: isolate silo $i$ within a round.} Expanding over rounds by the chain rule,
\begin{equation}
I(D_i ; Y \mid D_{-i}) = \sum_{t=1}^{T} I\big(D_i ; Y^{(t)} \,\big|\, Y^{(<t)}, D_{-i}\big).
\end{equation}
Fix a round $t$ and condition on $Y^{(<t)}$ and $D_{-i}$. The global model
$\theta^{(t-1)}$ is a deterministic function of $Y^{(<t)}$ through the aggregation
rule. Hence for every $j \ne i$ the update $\Delta_j^{(t)}$ is a deterministic
function of $\theta^{(t-1)}$ and $D_j$, and the noise $\xi_j^{(t)}$ is drawn
independently. Writing the region sum containing $i$ as
\begin{equation}
Y_{A(i)}^{(t)} = w_i \big(\mathrm{clip}(\Delta_i^{(t)}, C) + \xi_i^{(t)}\big) + R,
\end{equation}
the remainder $R$ is a function of $Y^{(<t)}$, $D_{-i}$ and noise independent of
$D_i$, and is therefore conditionally independent of $D_i$ given the conditioning
variables. Region sums excluding $i$ are conditionally independent of $D_i$ for
the same reason.

We emphasise that the argument conditions on the past rather than asserting
unconditional independence between $D_i$ and the updates of other silos. Across
rounds those updates do depend on $D_i$ through the shared global model, and
conditioning on $Y^{(<t)}$ is precisely what makes the step valid.

\emph{Step 4: bound one round, then compose.}
The conditional channel from $D_i$ to $Y_{A(i)}^{(t)}$ is therefore Gaussian with
covariance $\big(\sum_{j \in A(i)} w_j^2 \sigma_j^2\big) C^2 I$ and mean shifted by
$w_i \,\mathrm{clip}(\Delta_i^{(t)}, C)$. Replacing $D_i$ by any other dataset
moves the clipped update by at most $2C$ in norm, since each clipped update has
norm at most $C$, and hence moves the mean by at most $2 w_i C$. The divergence
between two Gaussians sharing a covariance and differing in mean by $\Delta\mu$ is
$\|\Delta\mu\|^2 / (2\varsigma^2)$ for covariance $\varsigma^2 I$, giving
\begin{equation}
\frac{(2 w_i C)^2}{2 \big(\sum_{j \in A(i)} w_j^2 \sigma_j^2\big) C^2}
= \frac{2 w_i^2}{\sum_{j \in A(i)} w_j^2 \sigma_j^2} = m_i(\sigma),
\end{equation}
so Lemma~\ref{lem:maxkl} bounds each round's term by $m_i(\sigma)$. Summing over
$T$ rounds and adding $\ell_i$ gives the second entry of the minimum. The first
entry holds because $I(p_i ; \hat{p}_i) \le H(p_i)$ for any $\hat{p}_i$.

Both the left-hand side and the divergence computation refer to replacement of the
entire dataset $D_i$, so a single adjacency notion applies throughout and the
bound is established at silo granularity directly.
\end{proof}

\section{Proof of Theorem~\ref{thm:allocation}}
\label{app:alloc}

The obstacle is that the objective in \eqref{eq:opt} is a maximum over regions
and therefore non-smooth. The standard remedy is to name that maximum and
constrain it, which turns the problem into a smooth convex program whose
optimality conditions can be read off directly. The solution then follows in
three steps: verify that strong duality applies, show that no region's
constraint is slack at the optimum, and confirm that the resulting budget
equation has a unique root.

Introduce a slack variable $K$ and rewrite \eqref{eq:opt} as
\begin{equation}
\min_{K, \{S_r > 0\}} K
\quad \text{s.t.} \quad
\frac{a W_r}{S_r} + \ell_r \le K \;\; \forall r,
\qquad \sum_r S_r \le U .
\label{eq:convex}
\end{equation}
The objective is linear and each constraint is convex, since $S \mapsto aW_r/S$ is
convex and decreasing on $S > 0$. Setting $S_r = U / (2 |\Reg|)$ for all $r$ is
strictly feasible for $K$ sufficiently large, so Slater's condition holds and the
Karush-Kuhn-Tucker conditions are necessary and sufficient
\cite{boyd2004convex}.

\emph{Optimality conditions.} The Lagrangian is
\begin{equation}
\mathcal{L} = K + \sum_r \mu_r \left( \frac{a W_r}{S_r} + \ell_r - K \right)
+ \lambda \left( \sum_r S_r - U \right),
\end{equation}
with $\mu_r \ge 0$ and $\lambda \ge 0$. Stationarity in $K$ gives $\sum_r \mu_r =
1$. Stationarity in $S_r$ gives $-\mu_r a W_r / S_r^2 + \lambda = 0$, hence $S_r =
\sqrt{\mu_r a W_r / \lambda}$.

\emph{Every constraint binds.} Every per-region constraint is active at the
optimum. Suppose one were inactive for some $r$. Complementary slackness then
forces $\mu_r = 0$, and stationarity in $S_r$ forces $\lambda = 0$, which in turn
forces $\mu_{r'} = 0$ for all $r'$ and contradicts $\sum_r \mu_r = 1$. Activity of
every constraint gives $a W_r / S_r + \ell_r = K^{\star}$ for all $r$, which
rearranges to \eqref{eq:alloc}.

\emph{Uniqueness.} Define $g(K) = \sum_r a W_r / (K - \ell_r)$ on $(\max_r \ell_r,
\infty)$. It is continuous and strictly decreasing there, tends to $+\infty$ as
$K$ approaches the left endpoint, and tends to $0$ as $K \to \infty$. By the
intermediate value theorem there is exactly one $K^{\star}$ with $g(K^{\star}) =
U$. Bisection on this interval converges at a geometric rate, so the allocation is
obtained to tolerance $\epsilon$ in $O(|\Reg| \log(1/\epsilon))$ operations.

\section{Proof of Corollary~\ref{cor:delta}}
\label{app:delta}

The comparison is between two budgets required to reach the same privacy target.
The optimal allocation gives each region exactly what that region needs, whereas
uniform allocation gives every region what the most exposed one needs. The ratio
of the two budgets is what \eqref{eq:delta} names.

Assume $\ell_r = \ell$ for all $r$ and write $c = a / (K - \ell)$ for a target $K
> \ell$. By \eqref{eq:alloc} the optimal allocation requires $U_{\mathrm{opt}} = c
\sum_r W_r$.

Uniform allocation applies a single scale $\sigma$, giving $S_r = \sigma^2 V_r$.
The constraint for region $r$ reads $a W_r / (\sigma^2 V_r) + \ell \le K$, that is,
$\sigma^2 \ge c \, W_r / V_r = c \rho_r$. Satisfying every region therefore
requires $\sigma^2 = c \, \rho_{\max}$, and the resulting budget is
$U_{\mathrm{uni}} = c \, \rho_{\max} \sum_r V_r$.

Using $W_r = \rho_r V_r$,
\begin{equation}
\frac{U_{\mathrm{opt}}}{U_{\mathrm{uni}}}
= \frac{\sum_r \rho_r V_r}{\rho_{\max} \sum_r V_r}
= \frac{\langle \rho \rangle_V}{\rho_{\max}},
\end{equation}
and $\delta = 1 - U_{\mathrm{opt}} / U_{\mathrm{uni}}$ gives \eqref{eq:delta}.
Since $\rho_r \le \rho_{\max}$ for every $r$, the ratio is at most one and $\delta
\ge 0$, with equality exactly when $\rho_r$ is constant across regions.

\end{document}